\documentstyle[prd,aps,epsf,epsfig]{revtex} 
\begin{document}
\draft
\twocolumn[\hsize\textwidth\columnwidth\hsize\csname@twocolumnfalse\endcsname
%
\title{Comment on "Percolation Thresholds in the Three-Dimensional Stick System}
%
\author{Z. N\'eda and R. Florian}
\address{ Babe\c{s}-Bolyai University, Dept. of Physics \\
          str. Kog\u{a}lniceanu 1, RO-3400 Cluj, Romania  }

\maketitle

\begin{abstract}

We discuss a classical mistake made in earlier publications on the stick 
percolation problem in 3D for generating
the right isotropic configuration of sticks. We explain the observed systematic
deviations from the excluded volume rule. New MC simulations considered by
us confirm nicely the applicability of the excluded volume theory.

\end{abstract}

]


The three-dimensional continuum percolation problem of permeable sticks
with the form of capped cylinders was considered by Monte Carlo simulations
in Ref. 1. The authors report the dependence of the percolation 
threshold on aspect ratio and on macroscopic anisotropy, discussing the results from the viewpoint of the excluded volume theory $^2$.

	Throughout the paper$^1$ the authors claim to obtain the isotropic 
distribution of the rods orientations by generating their $\theta$ and
$\varphi$ polar coordinates randomly with a uniform distribution
on the $[-\pi/2, \pi/2]$ and $[0, 2\pi]$ intervals, respectively. Following
their two-dimensional study$^3$ they define the measure of the macroscopic 
anisotropy of the system as:
\begin{equation}
P_{\parallel}/P_{\perp}=\sum_{i=1}^N \mid cos(\theta_i) \mid / \sum_{i=1}^N
[1-cos^2(\theta_i)]^{1/2}
\end{equation}
Proceeding however, in the way described above, the generated configurations 
will definitely not be isotropic ones, although their anisotropy constant
(1) will be one. It is easy to realize that the $z$ axes will be a privileged
one, and percolation in this direction reached easier than in the $y$ or
$x$ direction. In order to get the right isotropic distribution for the 
rods orientation, their endpoints must span uniformly the surface of a sphere. 
This can be achieved only by choosing the $\theta$ angle randomly with a weighted 
distribution and not a uniform one. From the surface element on the unit-sphere
($d\sigma= sin(\theta)\: d\theta \: d\varphi$)
it is immediate to realize that the weight-factor is governed by the
$sin(\theta)$ term.

The mistake made by the authors does not effect the $L<<r$ ($L$ the length of
the cylinder and $r$ itd radius) limit, considered by the authors to get 
confidence in their simulation data. However, when calculating the $\rho_c$
critical density at percolation and the $V_{ex}$ excluded volume of the
sticks
\begin{eqnarray}
& \rho_c=\frac{1}{V_{ex}} \\
& V_{ex}=(32 \pi /3) r^3+8 \pi Lr^2+4L^2r<sin(\gamma)>
\end{eqnarray}
they calculate the average of $sin(\gamma)$ ($\gamma$ the angle between two
randomly positioned sticks) for the right isotropic case, getting 
$<sin(\gamma)>=\pi/4$. Calculating $<sin(\gamma)>$  for their
"isotropic" configurations the result would be $<sin(\gamma)>=2/\pi$.
It is even more striking that in a following letter$^4$, confirming also
the excluded volume theory, the authors do observe the systematic deviation of
the Monte Carlo results$^1$ for the isotropic case respective to the 
excluded volume rule (Fig. 2 in Ref. 4), but they fail in explaining it.
In Ref. 4 the authors argue that the systematic deviation is due to the fact
that much smaller aspects ratios are required to get the right 
$r/L\rightarrow 0$ limit. The difference however, is obvious and in perfect
agreement with our previous affirmations. The real value of $<sin(\gamma)>$
for the "isotropic" Monte Carlo simulations$^1$ should be $2/\pi$, which is
approximately $1.24$ times smaller than the value used ($\pi/4$), and the 
systematic deviation$^4$ in Fig.2 is just of this order in the right direction.
The error in generating the right isotropic distribution is repeated in
a rapid publication$^5$, where the authors study by Monte Carlo methods
the cluster structure and conductivity of three-dimensional continuum
systems. The simulation data for the isotropic system$^1$ is used in a 
series of other papers$^{6-10}$, where some tables and comparison with analytical results should be reconsidered.

New MC simulation considered by generating the right isotropic
configuration confirm nicely the excluded volume theory.
We will discuss our new simulation datas and the simulation procedure
in a regular paper.

\begin{enumerate}
\item I. Balberg, N. Binenbaum and N. Wagner; Phys. Rev. Lett. {\bf 52}, 
1465 (1984)
\item I. Balberg, C.H. Anderson, S. Alexander and N. Wagner; Phys. Rev. B
{\bf 30}, 3933 (1984)
\item I. Balberg and N. Binenbaum; Phys. Rev. B {\bf 28}, 3799 (1983)
\item A.L.R. Bug, S.A. Safran and I. Webman; Phys. Rev. Lett. {\bf 54}, 
1412 (1985)
\item I. Balberg and N. Binenbaum; Phys. Rev. A {\bf 31}, 1222 (1985)
\item I. Balberg; Phys. Rev. B {\bf 31}, 4053 (1985)
\item I. Balberg; Phil. Mag. B {\bf 56}, 991 (1987)
\item I. Balberg and N. Binenbaum; Phys. Rev. A {\bf 35}, 5174 (1987)
\item D. Laria and F. Vericat; Phys. Rev. B {\bf 40}, 353 (1989)
\item A. Drory; Phys. Rev. E {\bf 54}, 5992 (1996)
\end{enumerate}

\end{document}